\documentclass[global,twocolumn]{svjour}
\usepackage{graphics}
\usepackage{epstopdf}
\begin{document}
\title{Switchable Magnetic Bottles and Field Gradients for Particle Traps}
\author{Manuel Vogel\inst{1,2} \and Gerhard Birkl\inst{1} \and Wolfgang Quint\inst{2,3} \and David von Lindenfels\inst{2,3} \and Marco Wiesel\inst{1,2,3}
}                     
\institute{Institut f\"ur Angewandte Physik, Technische Universit\"at Darmstadt, 64289 Darmstadt, Germany \and Helmholtz-Zentrum f\"ur Schwerionenforschung GSI, 64291 Darmstadt, Germany \and Physikalisches Institut, Ruprecht Karls-Universit\"at, 69120 Heidelberg, Germany}
\date{Received: date / Revised version: date}
%
\maketitle
\begin{abstract}
Versatile methods for the manipulation of individual quantum systems, 
such as confined particles, have become central elements in current developments in precision spectroscopy, frequency standards, quantum information processing, quantum simulation, and alike.
For atomic and some subatomic particles, both neutral and charged, a precise control of magnetic fields
is essential.
In this paper, we discuss possibilities for the creation of specific magnetic field configurations which find application in these areas.
In particular, we pursue the idea of a magnetic bottle which can be switched on and off by transition between the normal and the superconducting phase of a suitable material in cryogenic environments, for example in trap experiments in moderate magnetic fields. Methods for a fine-tuning of the magnetic field and its linear and quadratic components in a trap are presented together with possible applications.
\end{abstract}
\section{Introduction}
Traps for charged particles, in particular Paul- and Penning traps \cite{werth1,werth2,gho}, find application within precision spectroscopy from the radio-frequency to the x-ray regime of photon energies, within quantum state preparation, quantum simulation and possibly quantum computation, within anti-particle and anti-matter studies, and within experiments with non-neutral plasmas including Coulomb ion crystals. They provide techniques for particle confinement and cooling of the particle motion and hence are ideal tools for precision studies on long time scales. They also provide numerous manipulation techniques where the confining fields are used to create a specific influence on the confined particles which interrogates the properties of interest or creates desired states. 

Overviews of trap principles, realisations and applications have been given in \cite{werth1,werth2,gho}. Of fundamental importance for many applications is well-defined confinement, detailed knowledge of the ion motion in the trap, and in many cases motional cooling, which is true both for single-ion
experiments and confined ion clouds or plasmas. Corresponding treatments for Penning traps have for example been given in \cite{BRO86,gab89,john} and are based on homogeneous magnetic fields. 

\begin{figure}[h!] \begin{center}
\resizebox{\columnwidth}{!}{\includegraphics{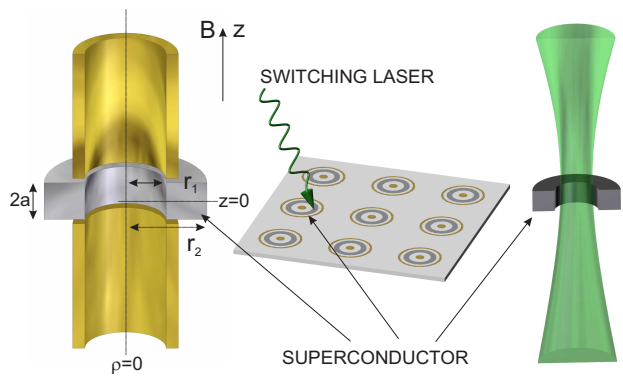} }
\caption{Application examples for (switchable) magnetic bottles in particle traps. Left: cylindrical Penning trap with the ring electrode forming a magnetic bottle. Centre: Array of planar Penning traps on a substrate with locally switchable magnetic bottle(s). Right: Optical dipole trap for neutral atoms with magnetic bottle.}
\label{var} \end{center}
\end{figure}

Also for neutral atoms, magnetic fields, sometimes in combination with optical 
fields, represent an important experimental ingredient for confining, cooling, and manipulating
atomic quantum systems. One prominent example for exciting applications is given by the realization and investigation of Bose-Einstein condensates \cite{top,a,a1} and quantum-degenerate fermionic matter in dilute atomic gases \cite{b}. A large variety of trap configurations, based on inhomogeneous magnetic fields, are used, including the 
magneto-optical trap (MOT) \cite{c}, a workhorse in many current experiments, conservative magnetic 
traps for pure magnetic confinement \cite{phil,iof}, and integrated and miniaturized trap configurations, typically 
summarized as 'atom chips' \cite{micro1,micro2,micro3,micro4}. 

Figure \ref{var} gives an impression of applications of (switchable) magnetic bottles, in Penning traps (left), in planar trap arrays with individual switching by a heating laser (centre), and in optical dipole traps where a (switchable) magnetic bottle can be superimposed to the optical configuration. In this paper, we present a class of methods based on the temporal control of specific magnetic field configurations which deviate from the homogeneous case and discuss experimental possibilities connected with those.

\section{Applications and Benefits of Field Gradients and Magnetic Bottles}
Magnetic field gradients find numerous applications, for example in magnetic mirrors \cite{mir}, magnetic bottle spectrometers \cite{magbot1,magbot2} or Zeeman slowers \cite{slow1,slow2}.
In connection with traps, the shaping of fields by introduction of gradients has two main aspects. One is that in a position-dependent field we may choose the absolute value of the field seen by the particles by according positioning. In presence of a field gradient, a shift of the position will result in a change of the absolute value of the magnetic field at the position of the ion and hence changes the value of the Zeeman splitting of spectral lines and so forth, which finds application for example in (Zeeman-tuned) level-crossing spectroscopy \cite{dem,ps,mag,mag2,bud,bud2,ch4}, magneto-optical traps \cite{c} and individual addressing of trapped particles via a position-dependent magnetic field \cite{ind}, particularly in quantum logic with confined particles \cite{mint}. 

The other important aspect is the geometry of the field itself which can give rise to desired effects, for example magnetic gradient forces on dipoles \cite{boj} as used in time orbiting potential (TOP) traps \cite{f}, in Ioffe-Pritchard traps \cite{iof}, and various magnetic microtraps \cite{micro1,micro2,micro3,micro4}. 
Quadratic distortions of the magnetic field in the shape of a so-called 
'magnetic bottle' are valuable to Penning-trap experiments, for example when the continuous Stern-Gerlach effect \cite{deh} is used for measurements of magnetic moments of unbound electrons \cite{vd87,hanneke}, unbound protons and antiprotons \cite{ulm}, and bound electrons in highly-charged ions \cite{her,haff,haff2,verd,ver,sturmSi,wag}, or when defined couplings amongst oscillatory degrees of freedom in the Penning trap are used for spectroscopic purposes \cite{sensors,njp}. Nested Penning traps \cite{nest} can benefit from introduced magnetic field gradients, regarding for example Penning-Ioffe traps as used in antihydrogen research \cite{anti}.

In the following, we will discuss the generation, measurement, and 
application of magnetic field gradients and in particular magnetic bottles to experiments with confined particles, and pursue the idea of a switchable magnetic bottle. 

\section{Magnetic Bottles}
We will speak of a magnetic bottle in terms of a field configuration with a quadratic component of the form
\begin{equation}
\label{bot}
\vec{B}(z,\rho)=B_2\left(\frac{2z^2-\rho^2}{2} \vec{e_z} - \rho z \vec{e_\rho} \right)
\end{equation}
where $\vec{e_z}$ is the axial symmetry axis and $\vec{e_{\rho}}$ is the radial unit vector. 
The parameter $B_2$ is a measure of the bottle strength as defined in the expansion of the absolute value of the magnetic field near the bottle centre 
\begin{equation}
\label{devel}
B_z(z,\rho)=B_0-2B_1z+B_2\left(z^2-\frac{1}{2}\rho^2 \right) + \dots
\end{equation}
where $B_0$ is the homogeneous part of the field and $B_1$ describes a gradient along $\vec{e_z}$. 

Such a field configuration is e.g. produced by a material of certain radially symmetric geometry when its magnetization is different from that of its surroundings.
For the sake of an analytical calculation, we assume this material to be an annular disc, i.e. a torus of rectangular cross section, with inner radius $r_1$, outer radius $r_2$ and a thickness of $2a$ in an external homogeneous magnetic field (see also figure \ref{var}, left), as they are used in several experiments \cite{her,haff,haff2,verd,ver,sturmSi,wag}. For simplicity, we will speak of a 'disc' from here on. In principle, the geometry may be chosen differently, however a mathematical description then becomes tedious, see for example \cite{verd2}. This would not change the validity of the arguments to follow. 

First, we are interested in the magnetic field contributions in the disc centre ($z=\rho=0$). They can be calculated by use of Maxwell's equations with the scalar magnetic potential $\Phi$ given by $\Delta \Phi = \vec{\nabla} \cdot \vec{M}$, where $\vec{M}$ is the magnetisation which we 
assume to be homogeneous throughout the disc and equal to $M_0$. 
The scalar magnetic potential written in terms of the magnetic surface charge density $\sigma_M=\vec{n} \cdot \vec{M}$ on the z-axis is thus given by
the integral over the closed surface $S$ \cite{jack}
\begin{equation}
\Phi(z)=\oint_S \frac{\sigma_M}{|\vec{x}-\vec{x'}|} da'
\end{equation}
which for our homogeneously magnetised disc is
\begin{equation}
\label{zwo}
M_0 \int_0^{2\pi} \!\!\! \int_{r_1}^{r_2} \!\!\! \frac{\rho'}{\sqrt{(z-a)^2+\rho'^2}} \!-\! \frac{\rho'}{\sqrt{(z+a)^2+\rho'^2}} d\phi' d\rho'.
\end{equation}
Defining for convenience $v_{\pm}^{(1,2)}\!\!=\!((z \pm a)^2 + r_{(1,2)}^2)^{1/2}$
this integration results in
\begin{equation}
\Phi(z)=2 M_0 \left( v_-^{(2)}\! -v_-^{(1)}\! -v_+^{(2)}\! + v_+^{(1)} \right).
\end{equation}
The magnetic field on the axis is given as the corresponding derivative of the magnetic potential, i.e. $H_z(z)=\mbox{d}\Phi / \mbox{d}z$ such that
\begin{equation}
H_z(z)=2 \pi M_0 \left(  \frac{z-a}{v_-^{(2)}} - \frac{z-a}{v_-^{(1)}} - \frac{z+a}{v_+^{(2)}} + \frac{z+a}{v_+^{(1)}} \right).
\end{equation}
Defining further $w_{(1,2)}\!=\!(a^2+r_{(1,2)}^2)^{1/2}$, using that in vacuum we have $\vec{B}=\mu_0 \vec{H}$, and performing a series expansion of the square roots 
yields the value of the contribution of the disc to the homogeneous field in the disc centre ($z=\rho=0$) to be
\begin{equation}
\label{null}
\Delta B_0=\frac{\mu_0M_0}{2} \left( \frac{2a}{w_1^2} - \frac{2a}{w_2^2} \right) = \frac{\mu_0M_0}{2} \left( \frac{2a}{a^2+r_{1}^2} - \frac{2a}{a^2+r_{2}^2} \right).
\end{equation}
The linear gradient $B_1$ is zero in the disc centre due to the symmetry of the field. The quadratic contribution (bottle strength) is found to be
\begin{equation}
\label{res1}
B_2=\frac{\mu_0M_0}{2} \left[ \frac{3}{w_1} \left( -\frac{a}{w_1^2}+\frac{a^3}{w_1^4} \right) 
+ \frac{3}{w_2} \left( \frac{a}{w_2^2} - \frac{a^3}{w_2^4} \right) \right],
\end{equation}
which can be simplified to
\begin{equation}
\label{res2}
B_2=\frac{3\mu_0 M_0}{2} \left( \frac{2ar_1^2}{(a^2+r_1^2)^{5/2}} -  \frac{2ar_2^2}{(a^2+r_2^2)^{5/2}}  \right).
\end{equation}
Corresponding off-centre solutions will be presented in section \ref{resid}. 

Assuming a given $M_0$, the maximum of $B_2$ for the annular disc geometry is reached when $r_2 \gg r_1$ and the ratio $a/r_1$ is chosen such that the Legendre polynomial $P_4(\cos \beta)=35/8 \cos^4 \beta - 30/8 \cos^2 \beta +3/8$ evaluated at $\beta=\tan^{-1}(a/r_1)$ vanishes, see the discussion in \cite{BRO86}. This puts up the condition $a/r_1 \approx 0.577$ for annular discs. Slightly higher values
of $B_2$ can be reached for geometries which more closely follow the conditions imposed by the behaviour of $P_4$, for details see again \cite{BRO86}.

\begin{figure}[h!] \begin{center}
\resizebox{\columnwidth}{!}{\includegraphics{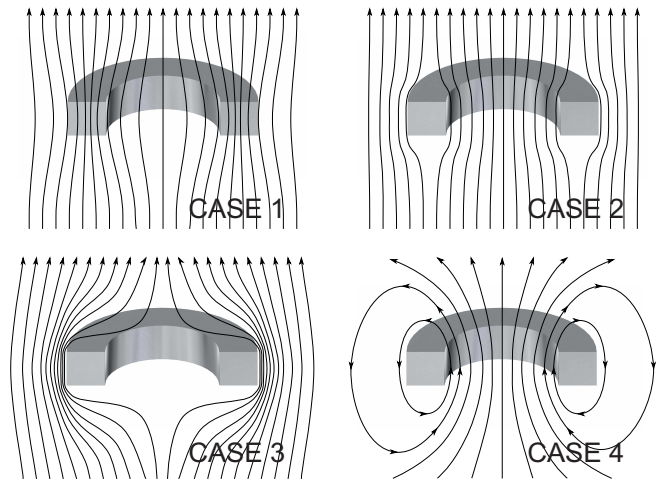} }
\caption{Schematic of the magnetic field configurations in cases 1 to 4, see text. The annular disc has been cut in half for better visibility.}
\label{four} \end{center}
\end{figure}

For the following, it 
is useful to visualise the resulting magnetic field lines. 
Concerning such a disc configuration and an external homogeneous magnetic field, we wish to distinguish four cases:
\begin{itemize}
\item {\bf case 1}: a ferromagnetic disc in an external magnetic field
\item {\bf case 2}: a disc is made superconducting within an already existing external magnetic field
\item {\bf case 3}: a magnetic field is established when a flux-free superconducting disc is already present 
\item {\bf case 4}: a flux-pervaded superconducting disc as in case 2 after the external magnetic field is removed again
\end{itemize}
Figure \ref{four} gives a qualitative picture of the field configurations in these cases. 
The field configurations have been simulated by use of the FEMM (Finite Element Method Magnetics) software package \cite{femm} and the artist's impression in figure \ref{four} is a quantitative exaggeration to make the effects visible to the eye.
We now discuss these cases in more detail.

\subsection{Ferromagnetic Bottle (case 1)}
\label{bott}
A magnetic bottle can be produced by a ferromagnetic annular disc which distorts the otherwise homogeneous magnetic field $B_0$ it is placed in. The strength of a magnetic bottle is given by its geometry and by its actual magnetisation in the external field, see equation (\ref{res2}). Many experiments which feature such bottles \cite{her,haff,haff2,verd,ver,sturmSi,wag} are performed in superconducting magnets of several Tesla strength, in this case the actual magnetisation $M_0$ is usually identical to the saturation magnetisation $M_s<B_0/\mu_0$ of the material. 
In the field line picture, the magnetization distribution of the disc distorts the external field such that field lines are attracted into the ferromagnetic material from the outside and from the inside of the annular disc (see also figure \ref{four}), i.e. the field strength in the centre ($z=\rho=0$) is smaller than $B_0$ by the amount $\Delta B_0$ given in equation (\ref{null}). Typical materials for disc electrodes to form a strong magnetic bottle come from the Nickel, Cobalt, Iron, CoFe, SmCo, NdFeB and AlNiCo families. The saturation field strengths $\mu_0M_s$ of specific high permeability iron alloys such as Hisat50 are as high as 2.44\,T. Using specialized geometries, magnetic bottles $B_2$ of up to about 400\,mT/mm$^2$ have been achieved \cite{dis,moos,hyb1,hyb2}. To illustrate this number, note that an axial particle shift by one millimetre changes the magnetic field seen by the particle by about one Tesla.

\subsection{Superconducting Bottle (case 2)}
Instead of a ferromagnet, the disc material may be chosen to be a superconductor. In principle, it acts like a perfect diamagnet, however has additional features. Initial
experiments to form a magnetic bottle from a superconducting current loop have been successful to create variable field gradients up to $B_2=2\mu$T/mm$^2$ \cite{vd,vd2}, however have not been pursued further
to experimental application. In its superconducting phase, the Meissner-Ochsenfeld effect \cite{meiss} provides that the interior of the superconducting material is field-free and hence distorts the outer magnetic field to form a magnetic bottle similarly to the ferromagnet, however of opposite sign and limited strength as will be discussed below.
Other than for the ferromagnetic disc from case 1 (as discussed in section \ref{bott}), for such a superconducting bottle we need to distinguish the three cases 2, 3 and 4 from above:

The external magnetic field penetrates the whole arrangement without any noteworthy distortion as long as the disc is normal-conducting. When it is cooled
below its critical temperature, the disc material expels the field from it, but the bore of the annulus still has magnetic flux passing through, albeit distorted in the shape of a magnetic bottle ({\bf case 2}), see also figure \ref{four}. This case is inverse to the ferromagnetic case (case 1) in the sense that field lines concentrate towards the central axis.
It is possible to 'trap' magnetic field lines when e.g. a circular disc is cooled radially from the outside to the inside and gradually becomes superconducting from the outside to the inside ('circle of frost'). Then all field lines which initially penetrated the whole disc are forced into the non-superconducting centre. If this centre is well-defined for example by different choice of material, the field shape and strength in this centre are well-defined as well, and it is possible to produce magnetic fields up to the critical field strength (see below) even in the absence of an outer magnet, simply by 'compression' of the Earth's magnetic field.

\begin{figure}[h!] \begin{center}
\resizebox{0.95\columnwidth}{!}{\includegraphics{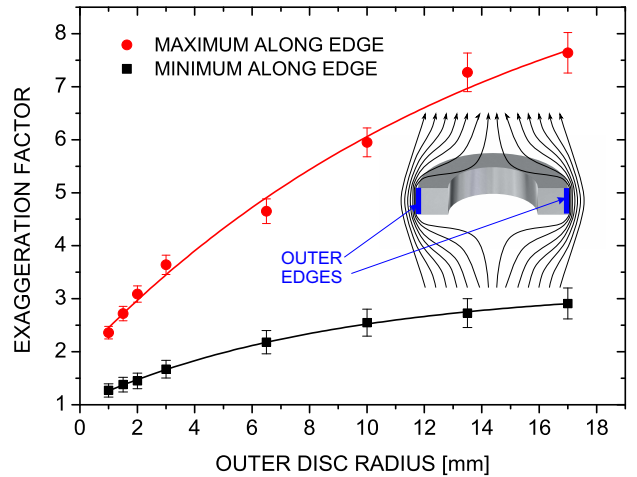} }
\caption{Magnetic field exaggeration factors $\eta$ as a function of the outer disc radius $r_2$ according to a FEMM simulation of case 3, the outer ring edges under discussion are marked in the inset (for details see text). Exponential saturation curves have been fit to guide the eye. They converge towards $\eta_{min}=3.2$ and $\eta_{max}=10.4$, respectively. The error bars reflect the finite resolution of the simulation.}
\label{disc} \end{center}
\end{figure}

When the disc is superconducting \textit{before} the external field is switched on, magnetic flux is expelled from the entire annular disc \textit{including} the bore ({\bf case 3}), see also figure \ref{four}. This is a consequence of the fact, that while in the superconducting state, the flux through the whole superconductor area cannot change since persistent currents in its outer surface shield the interior from the external flux exactly. 
This is true up to the point when residually penetrating fields (see section \ref{london}) exceed the critical field strength of the superconductor. For a given field, this determines the diameter of the inner disc region which stays superconducting. Figure \ref{disc} shows results from a FEMM \cite{femm} simulation, where the magnetic
field strength directly above the disc's outer surface (indicated in the inset of figure \ref{disc}) has been calculated in a given homogeneous magnetic field $B_0$ and for a given geometry. 
For a disc thickness of $2a=0.92$\,mm (the disc in \cite{her,haff,haff2,verd,ver,sturmSi,wag}), the field exaggeration factor $\eta$ (actual field strength directly above the surface measured in units of the ambient field strength $B_0$) is plotted as a function of the outer disc radius $r_2$. Note, that the choice of $r_1$ does not change these values. As can be seen, already for moderate disc
geometries, the exaggeration factor may reach values much above 1, such that outer parts of the disc lose superconductivity if the external field $B_0$ is chosen larger or of the order of $B_{c}/\eta$ where $B_c$ is the (lower) critical field strength of the superconductor at the given temperature, as discussed below. The induced persistent currents which maintain the magnetic flux within the superconductor represent a magnetic moment ${\vec{\mu}}$ which in presence of an outer magnetic field gradient ${\bf \nabla B}$ leads to a force $\vec{\mu} \cdot \nabla {\bf B}$.  
Simplifying the disc to a current loop of radius $R$ and picking a one-dimensional gradient in $z$-direction, the force $F_z$ induced by an outer magnetic field gradient $\partial B_0 / \partial z$ can be estimated by
\begin{equation}
F_z \approx \frac{2 \pi R^3 B_0}{\mu_0} \frac{\partial B_0}{\partial z} 
\end{equation}
which for $R=10$\,mm at a field of $B_0=1$\,T and a gradient of $\partial B_0/ \partial z=1$\,T/m would be 5\,N. For an annulus in a field with a gradient the real situation is more complicated, but the order of magnitude of the force remains true. This force can either be seen as mechanical stress on immovable parts or as a means to exert a force e.g. to remotely operate a mechanical switch or perform another motion in high magnetic field at low temperatures simply by heating or not heating e.g. with a laser. In a homogeneous outer magnetic field (as assumed so far), however, this effect does not occur as the gradient is zero. 

When the external field is switched off while the superconducting disc is flux-pervaded as in case 2, this results in magnetic flux trapping and the disc maintains the bottle field forever even in absence of an external field ({\bf case 4}). This may prove valuable to experiments which need a local magnetic bottle, but are otherwise sensitive to magnetic fields. One can hence 'load' the bottle initially, and then remove the outer field for the experiment to be performed. This is also an ideal tool to create arrays of permanent magnetic micro-bottles of identical strength, some of which can then individually be switched off again by short heating laser irradiation (as e.g. depicted in figure \ref{var}) and will remain switched off even after cooling back. This allows to manufacture complicated arrangements of planar (micro-)traps with and without magnetic bottles and may find application in quantum information processing.

\subsection{Minimum Dimensions}
\label{london}
In terms of a minimum possible size, the creation of a bottle from a superconductor is limited by finite values of the residual field penetration depth and of the coherence length. The London effect \cite{lon1,lon2} allows a certain non-vanishing magnetic field penetration into the superconductor. The penetrating magnetic field strength decays exponentially with the depth $d$ according to
\begin{equation}
B(d)=B_0 \exp \left(-\frac{d}{\lambda_L} \right)
\end{equation}
with a London penetration depth $\lambda_L$ on the scale of several tens of nanometres (for example 39\,nm for pure niobium, 37\,nm for lead \cite{kitt}). The same length scale is valid for the coherence length in typical superconductors which defines the smallest region in
which superconductivity can be achieved. For example, this length is 38\,nm for niobium and 83\,nm for lead \cite{kitt}. Hence, the minimum size of well-defined arrangements
is limited to micrometres and above, which does not restrict any of the applications to be discussed. The same is true for limits set by magnetic flux quantisation. The magnetic flux is quantized by nature \cite{kitt,doll}, however for structures of the present sizes, this quantization can be ignored due to the smallness of the quantum $\Phi = hc/2e \approx 2 \cdot 10^{-15}$\,Tm$^2$ \cite{kitt}. 

\subsection{Switching Superconductivity on and off}
Coming back to the case of a superconductor in an existing outer field ({\bf case 2}, which experimentally may be more prominent than the cases in which the outer field is switched), when the superconductor is slightly heated above its critical temperature, it loses superconductivity and thus the field distortion vanishes, i.e. the magnetic bottle is switched off. This process can be reversed back and forth within a short time and for an unlimited number of cycles. 
 
Type-I superconductors like lead, vanadium and tantalum have critical temperatures slightly above typical cryogenic temperatures like that of liquid helium and may thus be
interesting for applications. Their critical field strengths are limited to somewhat below 100\,mT, see table \ref{tab:1}. Also type-II superconductors like niobium are interesting for our purposes. They provide a complete Meissner effect up to a critical magnetic field strength $B_{c1}$. Above this field strength, the Meissner effect is incomplete ('Shubnikov phase' \cite{lon2}). Magnetic flux pinning by artificial material imperfections (doping) may help to shift the limits beyond $B_{c1}$, but this needs further study. Above a second critical field strength of $B_{c2}$, superconductivity is lost completely. Generally, the lower critical field strength depends on the actual temperature like \cite{lon2}
\begin{equation}
B_{c1}(T) \approx B_{c1}(0) \left( 1- \left( \frac{T}{T_c} \right)^2  \right),
\end{equation}
where the empirically found exponent for pure niobium is 2.13 instead of 2 \cite{finn}. The upper critical field strength has a temperature dependence given by \cite{bc2}
\begin{equation}
B_{c2}(T) \approx B_{c2}(0) \left( 1- \left( \frac{T}{T_c} \right)^2  \right)\left( 1+ \left( \frac{T}{T_c} \right)^2  \right)^{-1}.
\end{equation}
While the value of $B_{c2}$ can reach many Tesla, the lower critical field strength is typically only several tens to hundreds of mT. Table \ref{tab:1} lists the critical field strengths for several relevant superconductors.
\begin{table}
\begin{center}
\caption{Critical magnetic field strength of several relevant superconductors. Values are taken from \cite{lon2,finn,bc2}.}
\label{tab:1}       
\begin{tabular}{cccc}
\hline\noalign{\smallskip}
Superconductor & $B_{c1}(T=0)$ & $B_{c2}(T=0)$ & $T_c(B=0)$ \\
\noalign{\smallskip}\hline\noalign{\smallskip}
Pb & \multicolumn{2}{c}{81\,mT} & 7.2\,K \\
V & \multicolumn{2}{c}{102\,mT} & 5.3\,K \\
Ta & \multicolumn{2}{c}{83\,mT} & 4.5\,K \\
NbTi & 35\,mT & 10.3\,T & 9.3\,K \\
Nb$_3$Sn & 19\,mT & 24.5\,T & 18\,K \\
Nb & 180\,mT & 410\,mT & 9.25\,K \\
\noalign{\smallskip}\hline
\end{tabular}
\end{center}
\end{table}
Pure niobium has a critical field-free temperature of $T_c(B=0)$ of 9.25\,K \cite{finn} which is lowered in the presence of a magnetic field according to \cite{lon2}
\begin{equation}
T_c(B_0) \approx T_c(0) \sqrt{ 1- \left( \frac{B_0}{B_c}  \right)} .
\end{equation}
For example, at an ambient temperature of $T=4.2$\,K, the Meissner effect in pure niobium is complete up to a magnetic field strength of about 142\,mT. Hence, by choice of the 
applied magnetic field strength $B_0$ close to this value, the critical temperature $T_c(B_0)$ is close to the given ambient temperature of 4.2\,K. It is thus possible
to shift the critical temperature $T_c(B_0)$ arbitrarily close to the ambient temperature, such that deliberate heating to break the superconducting state (bottle 'off') can be achieved
with arbitrarily small amounts of energy. The same argument can be turned around, such that a variation of the magnetic field strength at a fixed temperature can be used to
switch the bottle on and off.

It is advantageous for this application that the heat capacity particularly of metals is very small at cryogenic temperatures, for pure niobium at 4.2\,K it is around 300\,$\mu$J/g$^{-1}$K$^{-1}$ \cite{niob}. Let us look at an example: when a disc of pure niobium with a weight of 10\,g at 4.2\,K is to be heated by 100\,mK to
lose superconductivity, the required energy is 300\,$\mu$J, i.e. a heating power of 10\,mW (which is feasible in a typical cryogenic setup) needs to be applied for 30\,ms. When the heating is switched off, the superconducting state is soon restored within the cryogenic surrounding and the magnetic bottle is switched on again. 
While heating is feasible e.g. with an electric current through an attached resistor within short times, the cooling back to superconducting temperature may take longer, depending strongly on the details of heat conduction and radiation in the chosen arrangement. However, also other means of heating may be favourable when the desired magnetic field is likely to be disturbed by the electric current used for heating. Such alternatives comprise heating with thermal radiation or lasers. Since time
scales for heating and re-cooling depend largely on the total heat capacity of the superconductor, these can be made very small when for example small-scale planar traps with thin films of superconductors are considered \cite{stefan,gold}. 

\section{Magnetic Field Gradients}
\label{mag}
Field gradients such as linear slopes and magnetic bottles are valuable tools for specific manipulation of confined particles. Applications are
the precise control of individual oscillation frequencies of ions by position-dependent confining fields, axial and radial positioning of particles in a trap, tailored coupling of individual oscillatory degrees of freedom, adjustability of Zeeman splittings 
in the magnetic field (for example for double-resonance spectroscopy as discussed in \cite{pra,pra2}), and the so-called 'continuous Stern-Gerlach effect' for the observation of spin transitions of free or bound electrons in a magnetic bottle \cite{werth1,haff2}. 

In the following, we will speak about confined ions, keeping in mind that the presented concepts also apply to electrons, antiprotons and anions when signs are changed appropriately. 

\subsection{Generation of a Linear Field Gradient}
We discuss three possibilities to produce a well-defined non-zero magnetic field gradient $B_1$ by electromagnetic or magnetostatic arrangements.

\subsubsection{Maxwell Arrangement}
\label{max}
\begin{figure}[h!]
\resizebox{0.75\columnwidth}{!}{\includegraphics{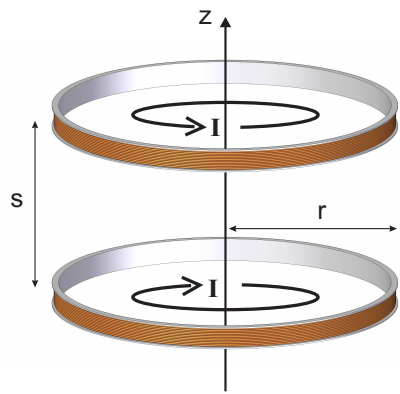} }
\caption{Maxwell arrangement of two circular solenoids with counter-propagating currents.}
\label{efig1}
\end{figure}

A pair of co-planar and coaxial loops with counter-propagating currents $I$ is a so-called 'Maxwell configuration', sometimes also
named 'anti-Helmholtz configuration'. This situation is depicted in figure \ref{efig1}. For loops with a radius $r$ and a 
separation $s$ along the $z$-axis, the axial magnetic field
is given by
\begin{equation}
B(z)=\frac{\mu_0 I r^2}{2 \left[ r^2+\left( \frac{s}{2}-z\right)^2 \right]^{3/2}} - \frac{\mu_0 I r^2}{2 \left[ r^2+\left( \frac{s}{2}+z\right)^2 \right]^{3/2}}.
\end{equation}
For $z$=0 (in the symmetry centre of the arrangement) the magnetic field $B(z=0)=0$. For reasons of symmetry, this is also true for all even expansion terms $B_2$, $B_4$,... at $z=0$.
The odd terms are given by
\begin{equation}
B_1=\frac{\partial B(z)}{\partial z}= \frac{3}{2}\mu_0Ir^2\frac{s}{\left[r^2+\frac{s^2}{4} \right]^{5/2}}
\end{equation}
and
\begin{equation}
B_3=\frac{\partial^3 B(z)}{\partial z^3}= \frac{15}{2}\mu_0Ir^2 \frac{s^3-3sr^2}{\left[r^2+\frac{s^2}{4} \right]^{9/2}}.
\end{equation}
The cubic term vanishes exactly if the separation $s$ is chosen to be $s=r\sqrt{3}$. In this case, all even terms are exactly zero for all values of $z$, and all odd terms except $B_1$ vanish or are
of order $z^5$ and higher, and thus can be considered negligible. This leaves a linear magnetic field gradient $B_1$ along the $z$-axis given by
\begin{equation}
B_1(z=0)=\frac{3^{3/2}}{2 \left( \frac{7}{4} \right)^{5/2}} \frac{N\mu_0I}{r^2}.
\end{equation}
Here, the factor $N$ describes a Maxwell arrangement of solenoids with $N$ windings each. This equation is a good
approximation as long as the geometry is close to the single-loop geometry. For a such an arrangement with $s=r\sqrt{3}$, and a choice of parameters $r$=30\,mm, $N$=100 windings at a current of $I$=1\,A, the resulting $B_1(z=0)$ is about 100$\mu$T/mm.

\subsubsection{Conical Solenoids}
\label{con}
\begin{figure}[h!]
\resizebox{0.75\columnwidth}{!}{\includegraphics{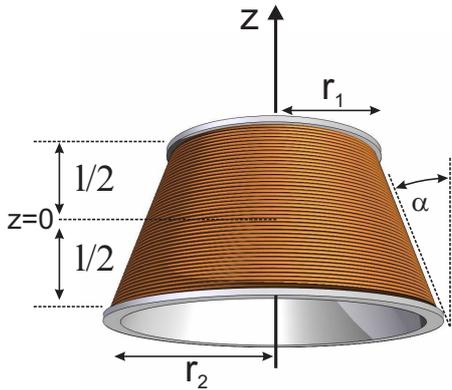}}
\caption{Conical solenoid with an opening angle $\alpha$ with respect to a cylindrical solenoid.}
\label{efig2}
\end{figure}

A linear magnetic field gradient can also be produced inside a solenoid with
conical geometry along the axis of the desired gradient. We assume a
solenoid of length $l$, radii $r_1$ and $r_2$ on the respective ends and $N$ windings with a current $I$. Let the centre of the cone be at $z=r=0$ such that
the ends are at $z=\pm l/2$ respectively, and $r_0=r(z=0)=(r_1+r_2)/2$, see figure \ref{efig2}. This geometry defines an opening angle $\alpha$ given by $\alpha=\tan^{-1}((r_2-r_1)/l)$. 
Starting with Biot-Savart's law for the on-axis field of $N$ current loops 
\begin{equation}
B_z(z)=\frac{IN\mu_0r^2}{2(r^2+z^2)^{3/2}}
\end{equation}
and integrating all contributions from $z=-l/2$ to $z=l/2$ along the conical geometry given by $r(z)=r_0+z \tan \alpha$ to the on-axis field in the centre $(z=0)$ yields an expression which can be expanded in terms of the opening angle $\alpha$. Neglecting higher-order contributions (since the opening angle is assumed small), one obtains the expression
\begin{equation}
B_1(z=0)=\frac{\partial B_z}{\partial z}(z=0) \approx \frac{12IN\mu_0l^2r_0 \tan \alpha}{(l^2+4r_0^2)^{5/2}}
\end{equation}
For $r_1=20$\,mm, $l$=50\,mm, $N=1000$,
$I=1$\,A and an opening angle of 5$^\circ$, the resulting linear field gradient $B_1$ is about $60 \mu$T/mm. Alternatively, as far as the magnetic field gradient on the symmetry axis and close to the centre is concerned, one could choose a cylindrical solenoid with a linearly changing number density of the
solenoid windings. Also configurations of co-axial and co-planar current loops with linearly increasing current as a function of axial loop position are possible, as well as co-axial and co-planar solenoid arrangements with same current but linearly changing numbers of windings. However, given the technical limitations of production and alignment, a single cone with constant winding density and only one current may be most promising as far as the achievable field definition is concerned. 

\subsubsection{Residual Gradients from a Magnetic Bottle}
\label{resid}
The on-axis ($\rho=0$, $z \neq 0$) magnetic field component of the bottle at a distance of $z$ to the ring centre is given by
\begin{eqnarray}
\label{q1}
B_z(z)&=&\frac{\mu_0M_0}{2} \mbox{\Big{(}}  \frac{z_+}{\sqrt{r_2^2+z_+^2}} - \frac{z_-}{\sqrt{r_2^2+z_-^2}} \\
&-& \frac{z_+}{\sqrt{r_1^2+z_+^2}} + \frac{z_-}{\sqrt{r_1^2+z_-^2}} \nonumber \mbox{\Big{)}} 
\end{eqnarray} 
where $z_{\pm}=z \pm a$.
The higher-order components of this field distortion are given by
\begin{equation}
\label{q2}
B_{(n)}(z)= \frac{\partial^n B_z(z)}{\partial z^n}
\end{equation}
and cause linear gradients $B_1 \neq 0$ and a residual magnetic bottle $B_2 \neq 0$ at positions along the central axis with $z \neq 0$. From equations (\ref{q1}) and (\ref{q2}), we obtain
\begin{equation}
B_1(z)=\frac{\partial B}{\partial z}(z)=-\frac{\mu_0M_0}{2} G_1 (z)
\end{equation}
where $G_1(z)$ is given by
\begin{eqnarray}
G_1(z)&=& \frac{r_1^2}{\left(r_1^2+(a+z)^2\right)^{3/2}}-\frac{r_1^2}{\left(r_1^2+(a-z)^2\right)^{3/2}}  \nonumber \\
  &+& \frac{r_2^2}{\left(r_2^2+(a-z)^2\right)^{3/2}}-\frac{r_2^2}{\left(r_2^2+(a+z)^2\right)^{3/2}}.
\end{eqnarray}
As discussed earlier, for $z=0$ the linear gradient $B_1$ vanishes as the geometry factor $G_1(z=0)$ vanishes.

The residual magnetic bottle strength $B_2$ at the position $z$ along the central axis is given by the second derivative of $B$ with respect to $z$
\begin{equation}
B_2(z)= \frac{\partial^2 B}{\partial z^2}(z)=-\frac{\mu_0M_0}{2} G_2(z) 
\end{equation}
where $G_2(z)$ is given by
\begin{eqnarray}
G_2(z)&=& -\frac{3r_1^2 (a+z)}{\left(r_1^2+(a+z)^2\right)^{5/2}}-\frac{3r_1^2 (a-z)}{\left(r_1^2+(a-z)^2\right)^{5/2}}  \nonumber \\
  &+& \frac{3r_2^2 (a-z)}{\left(r_2^2+(a-z)^2\right)^{5/2}}+\frac{3r_2^2 (a+z)}{\left(r_2^2+(a+z)^2\right)^{5/2}}.
\end{eqnarray}
For $z=0$ we obtain the expression at the ring centre as given earlier by equation (\ref{res2}). Higher-order contributions can be obtained from further application of equation (\ref{q2}) for $n=3,4$ and so forth, these however are not of present interest.

The residual magnetic bottle strength $B_2(z \neq 0)$ at a position along the axis is generally not negligible and is commonly regarded as a disadvantage, since it introduces an additional dependence of the field on the radial coordinate, as can be seen in equation (\ref{devel}). 

For example, in experiments for microwave spectros\-copy of electrons bound in hydrogen-like ions, a ferromagnetic bottle of strength $B_2=10$mT/mm$^2$ based on a nickel ring has been used. An initial measurement of the magnetic moment of the electron bound in hydrogen-like carbon $^{12}$C$^{5+}$ \cite{her} was limited due to the permanent presence of a magnetic bottle to a relative accuracy of $1 \cdot 10^{-6}$.
To avoid this, successor experiments have located this part of the measurement away from the magnetic bottle ('double-trap technique') which helped to shift the relative accuracy to $10^{-9}$ and better \cite{haff,haff2,verd,ver,sturmSi,wag}. In this case, a switchable magnetic bottle may have been desirable, as it would have avoided the influence of the permanent magnetic bottle, which has created a residual field gradient of $B_1=60\mu$T/mm, a residual $B_2$=4\,$\mu$T/mm$^2$ and a fourth-order contribution of $B_4$=11\,nT/mm$^4$ at the position of the precision frequency measurement, $z$=28\,mm away from the ring centre along the axis.
Superconducting magnetic bottles are restricted to fields $B_0 < B_c$ about one order of magnitude smaller than the ones used here, however, similar magnetic bottle strengths $B_2$ are possible. Ion transport between traps becomes unnecessary and the related efforts and uncertainties \cite{haff2} are avoided. Penning trap experiments in low magnetic fields
are mainly subject to two technical issues:
\begin{itemize}
\item Low magnetic fields lead to low ion oscillation frequencies such that electronic noise, particularly $1/f$-noise, may be significant.
\item Low magnetic fields require small electrostatic trapping potentials $U_0$ which may become subjected
to surface effects (patch effects) when voltages approach the level of few Volts. 
\end{itemize}
Hence, electronic detection of the ion motions tends to become more difficult in low fields. This can be an issue for experiments like the ones of Stern-Gerlach type \cite{her,haff,haff2,verd,ver,sturmSi,wag}, but not necessarily for optical experiments like the ones discussed in \cite{sensors,njp,pra,pra2}.

\subsection{How to Measure Magnetic Field Gradients}
The value of a linear magnetic field gradient $B_1$ and of a magnetic bottle strength $B_2$ (and higher orders) can be determined by a position-dependent measurement of the effective magnetic field strength. 
This can for example be achieved in a measurement of the ion cyclotron frequency $\omega_c=qB/m$ as a function of the axial (and/or radial) ion position, where $q$ is the ion charge and $m$ is its mass. 

Alternatively, the field distortion can be derived from a position-dependent measurement of the Zeeman splitting in a known ion or any other property that depends on the absolute value of the magnetic field in a predictable way.

In a Penning trap, the cyclotron frequency is perturbed by the presence of the axial trapping potential $U_0$ which aside from the perturbed cyclotron oscillation at $\omega_+$ evokes an axial oscillation at $\omega_z$ and a drift motion in the $\vec{E} \times \vec{B}$ field at $\omega_-$ \cite{werth1,gho,BRO86,gab89}. 
The common way to determine $\omega_c$ is via the 'invariance theorem' $\omega_c^2=\omega_+^2+\omega_z^2+\omega_-^2$, which is robust against typical small experimental imperfections like tilts and ellipticities of the electrostatic fields with respect to the magnetic field \cite{inv1,inv2}. In the presence of a magnetic bottle, the invariance theorem does not strictly hold, but there is a second-order effect of the shifted individual frequencies which, measured by the shift $\Delta \omega_+$ of the perturbed cyclotron frequency, is given by
\begin{equation}
\frac{\Delta \omega_c}{\omega_c}\approx \left(1+\frac{1}{2} \left( \frac{\omega_+-\omega_-}{\omega_z} \right)^2 \right) \left( \frac{\Delta \omega_+}{\omega_+} \right)^2,
\end{equation}
where the shift $\Delta \omega_+$ resulting from finite motional energies $E_{\pm}$ and $E_z$ in the magnetic bottle is given by
\begin{equation}
\Delta \omega_+=\frac{\omega_+}{m \omega_z^2} \frac{B_2}{B_0} \left(  E_z - \frac{\omega_z^2}{\omega_+^2} E_+ +2 E_- \right).
\end{equation}
This results in $\Delta \omega_c / \omega_c \approx 10^{-13}$ for experimental conditions as typically present e.g. in \cite{her,haff,haff2,verd,ver} and may hence be ignored in these cases. It may, however, become relevant for stronger magnetic bottles and increased experimental accuracies which can be obtained by use of novel measurement techniques \cite{sturm}. 
 
To position the ion(s), the trap can be made electrically asymmetric by an additional small voltage $U_a$ symmetrically across the trap endcaps.  
In a cylindrical Penning trap, this adds a small uniform axial electric field $E_z=c_1U_a/2z_0$ close to the trap centre and shifts the centre position of the axial oscillation by an amount $\Delta z$ given by \cite{gab84}
\begin{equation}
\label{axsh}
\Delta z= \frac{d^2}{2z_0}  \frac{c_1}{C_2} \frac{U_a}{U_0},
\end{equation}
where $c_1$ and $C_2$ are dimensionless geometry coefficients of order unity \cite{BRO86,gab89,gab84} and $d^2=z_0^2/2+\rho_0^2/4$ is the effective trap size resulting from its inner diameter $\rho_0$ and axial size $z_0$. Note, that a non-zero value of $c_1$ shifts the axial frequency $\omega_z=(qU_0/md^2)^{1/2}$ by an amount
\begin{equation}
\label{wsh}
\frac{\Delta \omega_z}{\omega_z}=-\frac{3}{4} \frac{d^4}{z_0^4} c_1 c_3 \left( \frac{U_a}{U_0} \right)^2,
\end{equation}
where $c_3$ is a dimensionless geometry coefficient which for common trap geometries obeys the relation $c_1+c_3 \approx 1$ \cite{BRO86,gab84}. This shift may need to be considered when the magnetic field strength is deduced from the measured oscillation frequencies, depending on the desired accuracy. 
The range of possible centre shifts $\Delta z$ is mainly limited by the potential asymmetry $U_a$ giving rise to additional anharmonicities. The leading contribution is given as a change of the $C_4$-coefficient \cite{gab84,gab89}
\begin{equation}
C_4 \rightarrow C_4 - \frac{5}{4} c_3^2 \left(  \frac{U_a}{U_0} \right)^2.
\end{equation}
The general recipe for the calculation of such potential influences is given in \cite{BRO86}. A corresponding measurement of the magnetic bottle strength $B_2$ for the trap in \cite{her,haff,haff2,verd,ver} has been performed with single ions \cite{haff2} and yielded results for the actual $B_2$ in good agreement with the prediction by equation (\ref{res2}). The method can similarly be applied
with different trap geometries like planar traps \cite{stefan,gold} depending on the possibility of ion positioning.

Due to the finite motional amplitude of the ion, the result of a field measurement is always a time-averaged value.
This is identical to the effective field at the centre of the motion as long as only non-zero
$B_0$ and $B_1$-terms are present. For a non-zero term $B_2$, the time-averaged
field will be higher if $B_2 >0$ and lower if $B_2 <0$ due to the quadratic behaviour in equation (\ref{bot}). It is then necessary to cool the ion to 
oscillation amplitudes much smaller than the range of the shift. This is particularly easy with highly-charged ions due to the strong binding in the trapping potential which allows  axial amplitudes of the order of 1\,$\mu$m and smaller even at liquid helium temperatures. For optical spectroscopy this also means that the Lamb-Dicke regime can be reached
without sophisticated cooling methods simply by resistive cooling around liquid helium temperature \cite{njp}.
Similar measurements are possible also with coherent ensembles of ions as long as the space charge effect \cite{yu,win} is negligible. 

\subsection{Optical Transition Rate Measurements}
\label{tra}
An artificial electric asymmetry $U_a$ of a trap as presented in the previous section can be combined with a magnetic bottle to allow a measurement of optical transition rates, using a purely electronic measurement, i.e. without optical detection \cite{sensors}. 
This can find application in spectroscopy experiments like the ones discussed in \cite{sensors,njp}, in \cite{pra,pra2} and \cite{spec}, and would be an interesting addition to the Stern-Gerlach experiments of the kind \cite{her,haff,haff2,verd,ver,sturmSi,wag} when an ion with a suitable optical transition is considered.
We assume an ion in the centre of a magnetic bottle which is
laser-cooled along the trap axis. The light pressure shifts the ion along the axis out of the trap centre, which in the presence of a magnetic bottle, results in a shift of the radial oscillation frequencies and can be measured non-destructively. An axially asymmetric
trapping potential is used to restore the ion position which is seen as a vanishing shift of the radial oscillation frequency. This potential asymmetry directly
determines the desired value of the optical transition rate $\Gamma$. To see this, the shift $\Delta z$ of the axial ion position is obtained from balancing the 
force $F_l$ of the laser (for simplicity driving the transition in saturation) with the restoring force $F_e$ due to the electrostatic trapping potential, i.e. $F_l=F_e$ which reads
\begin{equation}
\frac{\hbar \Delta\omega \Gamma}{2c} = \frac{qC_2U_0 \Delta z}{d^2}
\end{equation}
and results in
\begin{equation}
\label{z2}
\Delta z=\frac{\hbar \Delta \omega \Gamma d^2}{2cqC_2U_0},
\end{equation}  
where $\Delta \omega$ is the laser detuning. This shift of the axial position (and thus the shift of the measured radial frequencies) can be restored if the electrostatic trapping
potential is made asymmetric by an additional voltage $U_A$ across the endcaps which shifts the axial position by an amount
given by equation (\ref{axsh}). Now we balance the two shifts $\Delta z$ from equations (\ref{z2}) and (\ref{axsh}) and solve for the desired optical transition rate to get
\begin{equation}
\label{ff}
\Gamma = \frac{cqc_1}{\hbar \Delta \omega z_0} U_a.
\end{equation}
Comfortably, on the right-hand side, we find only constants and well-controllable parameters.
Also, equation (\ref{ff}) is independent of the electric trapping potential $U_0$ such that we are free to choose the axial oscillation frequency to
take a convenient value. The obtainable precision depends mainly on the knowledge of the trap parameters $c_1$ and $z_0$ and on the knowledge and temporal stability of the voltage asymmetry $U_a$. Typical relative precisions of trap geometry parameters are on the $10^{-3}$ to $10^{-4}$ scale. By gauging with a known ion this can be exceeded to the extent the ion and laser parameters allow for. Finally, a technical limit is set by the voltage $U_a$ which normally cannot be controlled and known better than to some $10^{-6}$. Nonetheless, this scheme may prove valuable, as in many situations appropriate lasers exist but light collection and optical detection is tedious or impossible, e.g. in the XUV and infrared regimes. This problem is circumvented here since the scheme is independent from optical detection.

\subsection{Magnetic Field Control}
For a number of applications, the absolute value of the magnetic field needs to be set with high accuracy. For applications like microwave spectroscopy of Zeeman transitions (see for example \cite{pra,pra2}) one additionally
desires the possibility to scan over a certain range of the field strength in a fast, reproducible and uncomplicated way. 
Coarse tuning of the magnetic field strength is possible with the current of the main magnet solenoids, which typically can be done with an accuracy of much better than one percent. At a field strength of some Tesla, this leaves a region of order 10 mT for fine-tuning. 

Such fine-tuning may be achieved for example by a small independent solenoid 
around the region of interest. This can be a Helmholtz-type arrangement as depicted in figure \ref{efig1} with the currents
applied in the same sense of rotation. With a typical
current of a few Amperes, the range of several tens of mT can be covered with a resolution of 0.1\% or better, limited by the accuracy of the current source. Hence, the total 
magnetic field strength can be set with an accuracy of the order of 10\,$\mu$T.
For further fine-tuning of this value and a scan across a certain region of field strengths, one can employ the position dependence
of the effective field strength in the presence of non-zero field gradients $B_1$ and/or $B_2$
in the same sense as discussed in section \ref{tra}. For the example of a Penning trap, the resulting axial shift $\Delta z$ is typically of the order of 1\,mm/V.
Combining this with equation (\ref{devel}) and neglecting the minor contribution from terms higher than $B_1$ results in a voltage-dependent 
magnetic field strength contribution given by
\begin{equation}
\Delta B = - B_1 \frac{z_0d^2C_1}{C_2U_0} U_a
\end{equation}
which can cover several hundreds of $\mu$T. The sign of $\Delta B$ can be chosen by the sign of $U_a$. Assuming a typical commercially available 
accuracy of $10^{-5}$ for the voltage $U_a$, the corresponding accuracy of $\Delta B$ is of order 1\,nT, which
represents a relative accuracy on the ppb level. This is comparable to the short-term field fluctuations of a typical superconducting magnet, to the
accuracy to which ion oscillation frequencies can be measured electronically, and also to the natural fluctuation of the Earth's magnetic field within a time of about one hour \cite{stacey}.

Calibration of
the solenoid current and electrode voltage settings to the absolute value of the obtained magnetic field strength can be achieved by an electronic measurement of the
cyclotron frequency of a well-known test ion like e.g. $^{12}$C$^{5+}$ in that field. 

Apart from applications in double-resonance \cite{pra,pra2} and radio-frequency spectroscopy \cite{haff2,sensors,njp}, such magnetic field control may also be used 
for level crossing spectroscopy, which has extensively been used to investigate properties of neutral atoms and singly charged ions (including molecular ions) 
even before the invention of lasers. An overview of applied techniques and performed measurements can be found in \cite{dem,ps}. 
Typically, particles are confined in gas cells or traps and a variable 
homogeneous magnetic (or electric) field is applied, see for example \cite{mag,mag2,bud,bud2}. Laser light is used to excite an electronic 
transition and the corresponding fluorescence is observed as a function of the
external magnetic field strength. When a level crossing of the observed levels with other field-dependent levels occurs, e.g. Zeeman sublevels of the
fine or hyperfine structure, the fluorescence is increased and thereby the corresponding value of the magnetic field is determined. For atoms and
singly charged ions, this value is typically of order mT and can thus be readily produced by electromagnets. Level crossing spectroscopy allows access
to the polarisability of states and to the magnetic dipole and electric quadrupole interaction constants as well as 
to the electronic Land\'e-factor $g_J$ \cite{dem,sva}.
When highly charged ions are considered, level crossings may occur for values of the magnetic field outside of the domain of non-superconducting magnets. 
It is then necessary to control magnetic fields of order Tesla with accuracies beyond the ppm regime and to readily perform well-defined scans over a certain
region of the field strength, as should be possible with the methods discussed in this section.

\section{Discussion}
We have seen that magnetic field gradients find numerous applications with confined particles, which is particularly obvious for linear gradients and magnetic bottles. It appears feasible to implement switchable magnetic bottles in cyrogenic experiments by choice of superconductor materials. Switching is possible on time scales much shorter than typical confinement times, since the amount of energy needed can be made arbitrarily small by appropriate choice of the temperature and/or magnetic field strength. This is even more the case when superconducting structures with small heat capacities are considered, like in planar traps or miniature optical dipole traps. Both in the 'on' and 'off' states, the magnetic field configuration is well-defined and the strength of the bottle does not change other than between zero and its given value. The critical field strengths of presently available superconductors is small, which limits the application to low magnetic fields of the order of hundred mT. In radio-frequency, magnetic, or optical traps, however, this is not a limitation for particle confinement and a switchable magnetic bottle can be superimposed to the confining configuration. Exploiting the specific properties of superconductors in external fields, we have seen that it possible to create magnetic bottles in otherwise field-free regions by magnetic flux trapping, allowing e.g. complex arrangements of magnetic micro-bottles.
It has been shown that well-defined magnetic field gradients can be produced, measured and used for various experimental techniques in spectroscopy such as laser-microwave double-resonance spectroscopy \cite{pra,pra2}, level crossing spectroscopy \cite{mag,mag2,bud,bud2}, and microwave and radio-frequency spectroscopy using the continuous Stern-Gerlach effect \cite{werth1,werth2,haff,haff2,verd}. Apart from this, they are the central prerequisite of several trap-specific spectroscopy schemes as have been discussed in detail in \cite{sensors,njp}. 

\begin{acknowledgement}
We dedicate this paper to the memory of Wolfgang Paul as an outstanding 
scientist, teacher, and founding father of this exciting field of 
physics. G.B. and W.Q. in addition would like to commemorate the late 
Herbert Walther for his important contributions to the field and for his 
enthusiasm. This work has been realised 
within the ARTEMIS project (AsymmetRic Trap for the measurement of Electron Magnetic moments in IonS). The authors 
would like to thank the rest of the ARTEMIS team for helpful 
discussions. This work has been supported financially in part by DFG 
(Grants No. VO1707/1-2 and No. BI 647/4-1) and GSI. D.L. is supported by 
IMPRS-QD Heidelberg.
\end{acknowledgement}
%


\begin{thebibliography}{}
\bibitem{werth1} G. Werth, V.N. Gheorghe and F.G. Major, \textit{Charged Particle Traps}, Springer, Heidelberg, 2005 
\bibitem{werth2} G. Werth, V.N. Gheorghe and F.G. Major, \textit{Charged Particle Traps II}, Springer, Heidelberg, 2009
\bibitem{gho} P. Ghosh, \textit{Ion Traps}, Oxford University Press, Oxford (1995) 
\bibitem{BRO86} L.S. Brown and G. Gabrielse, Rev. Mod. Phys. {\bf 58}, 233 (1986)
\bibitem{gab89} G. Gabrielse, L. Haarsma and S.L. Rolston, Int. J. Mass Spectr. Ion Proc. {\bf 88}, 319 (1989)
\bibitem{john} J.J. Bollinger, D.J. Wineland, and D.H.E. Dubin, Phys. Plasmas {\bf 1}, 1403-1414 (1994) and references therein
\bibitem{top} M.H. Anderson, J.R. Ensher, M.R. Matthews, C.E. Wieman and E.A. Cornell, Science {\bf 269}, 198 (1995)
\bibitem{a} K.B. Davis et al., Phys. Rev. Lett. {\bf 75}, 3969 (1995)
\bibitem{a1} M.-O. Mewes et al., Phys. Rev. Lett. {\bf 77}, 416 (1996)
\bibitem{b} B. de Marco and D.S. Jin, Science {\bf 285}, 1703 (1999)
\bibitem{c} S. Chu, Rev. Mod. Phys. {\bf 70}, 685 (1998) 
\bibitem{phil} W.D. Phillips, Rev. Mod. Phys. {\bf 70}, 721 (1998)
\bibitem{iof} D. Pritchard, Phys. Rev. Lett. {\bf 51}, 1336 (1983) 
\bibitem{micro1} E.A. Hinds and I.G. Hughes, J. Phys. D: Appl. Phys. {\bf 32}, R119 (1995)
\bibitem{micro2} J. Reichel, Appl. Phys. B {\bf 75}, 469 (2002)
\bibitem{micro3} R. Folman et al., Adv. At. Mol. Opt. Phys. {\bf 48}, 263 (2002)
\bibitem{micro4} J. Fort\'agh and C. Zimmermann, Rev. Mod. Phys. {\bf 79}, 235 (2007)
\bibitem{mir} S. Granger and G.W. Ford, Phys. Rev. Lett. {\bf 28}, 1479 (1972)
\bibitem{magbot1} T. Tsuboi et al., Rev. Sci. Inst. {\bf 59}, 1357 (1988)
\bibitem{magbot2} H. Handschuh, G. Gantef\"ohr and W. Eberhardt, Rev. Sci. Inst. {\bf 66}, 3838 (1995)
\bibitem{slow1} C.J. Dedman et al., Rev. Sci. Inst. {\bf 75}, 5136 (2004)
\bibitem{slow2} M.A. Joffe, W. Ketterle, A. Martin, and D.E. Pritchard, J. Opt. Soc. Am. B {\bf 10}, 2257 (1993)
\bibitem{dem} W. Demtr\"oder, \textit{Laser Physics}, 3rd edition, Springer, Heidelberg (2003)
\bibitem{ps} P. Hannaford, Physica Scripta {\bf T70}, 117 (1997)
\bibitem{mag} S. Rydberg and S. Svanberg, Physica Scripta. {\bf 5}, 209 (1972)
\bibitem{mag2} J. Alnis, K. Blushs, M. Auzinsh, S. Kennedy, N. Shafer-Ray and E.R.I. Abraham, J. Phys. B {\bf 36}, 1161 (2003) 
\bibitem{bud} B. Budick, S. Marcus and R. Novick, Phys. Rev. {\bf 140}, A1041 (1965) 
\bibitem{bud2} W. Hogervorst and S. Svanberg, Physica Scripta. {\bf 12}, 67 (1975)
\bibitem{ch4} A.C. Luntz and R.G. Brewer, J. Chem. Phys. {\bf 53}, 3380 (1970)
\bibitem{ind} S.X. Wang, J. Labaziewicz, Y. Ge, R. Shewmon and I.L. Chuang, Appl. Phys. Lett. {\bf 94}, 094103 (2009)
\bibitem{mint} F. Mintert and C. Wunderlich, Phys. Rev. Lett. {\bf 87}, 257004 (2001)
\bibitem{boj} T.H. Boyer, Americal Journal of Physics {\bf 56}, 688 (1988)
\bibitem{f} W. Petrich et al., Phys. Rev. Lett. {\bf 74}, 3352 (1995)
\bibitem{deh} H. Dehmelt, Proc. Natl. Acad. Sci. USA. {\bf 83} 2291 (1986)
\bibitem{vd87} R.S. Van Dyck, P.B. Schwinberg and H.G. Dehmelt, Phys. Rev. Lett. {\bf 59}, 26 (1987).
\bibitem{hanneke} D. Hanneke, S. Fogwell and G. Gabrielse, Phys. Rev. Lett. \textbf{100}, 120801 (2008).
\bibitem{ulm} S. Ulmer et al., Phys. Rev. Lett. {\bf 106}, 253001 (2011)
\bibitem{her} N. Hermanspahn et al., Phys. Rev. Lett. \textbf{84}, 427 (2000)
\bibitem{haff} H. H\"affner et al., Phys. Rev. Lett. {\bf 85}, 5308 (2000)
\bibitem{haff2} H. H\"affner et al., Eur. Phys. J. D \textbf{22}, 163 (2003)
\bibitem{verd} J. Verd\'{u} et al., Phys. Rev. Lett. {\bf 92}, 093002 (2004)
\bibitem{ver} J. Verd\'{u} et al., Physica Scripta {\bf T112}, 68 (2004)  
\bibitem{sturmSi} S. Sturm et al., Phys. Rev. Lett. \textbf{107}, 023002 (2011)
\bibitem{wag} A. Wagner et al., Phys. Rev. Lett {\bf 110}, 033003 (2013)
\bibitem{sensors} M. Vogel, W. Quint and W. N\"ortersh\"auser, Sensors {\bf 10}, 2169 (2010)
\bibitem{njp} M. Vogel and W. Quint, New J. Phys. {\bf 11}, 013024 (2009) 
\bibitem{nest} G. Gabrielse, S.L. Rolston, L. Haarsma, and W. Kells, Phys. Lett. A {\bf 129}, 38 (1988)
\bibitem{anti} G. Gabrielse et al., Phys. Rev. Lett. {\bf 100}, 113001 (2008)
\bibitem{verd2} J. Verd\'u et al., New Journal of Physics \textbf{10}, 103009 (2008)
\bibitem{jack} J.D. Jackson, \textit{Classical Electrodynamics}, 3rd edition, John Wiley and Sons (1998).
\bibitem{femm} D.C. Meeker, Finite Element Method Magnetics, Version 4.0.1 (03Dec2006 Build), http://www.femm.info
\bibitem{dis} J. DiSciacca and G. Gabrielse, Phys. Rev. Lett. {\bf 108}, 153001 (2012)
\bibitem{moos} A. Mooser et al., Phys. Rev. Lett. {\bf 110}, 140405 (2013)
\bibitem{hyb1} J. Verd\'u et al., AIP Conference Proceedings {\bf 796},  260 (2005)
\bibitem{hyb2} J. Verd\'u et al., New Journal of Physics {\bf 10}, 103009 (2008)
\bibitem{vd} P.B. Schwinberg and R.S. Van Dyck, Jr., Bull. Am. Phys. Soc. {\bf 27}, 889 (1982)
\bibitem{vd2} R.S. Van Dyck Jr., F.L. Moore, D.L. Farnham, P.B. Schwinberg, Rev. Sci. Inst. \textbf{57}, 593 (1986)
\bibitem{meiss} W. Meissner and R. Ochsenfeld, Naturwissenschaften {\bf 21} 787 (1933)
\bibitem{lon1} F. London and H. London, Proc. Roy. Soc. (London) {\bf A149}, 866 (1935) 
\bibitem{lon2} M. Tinkham \textit{Introduction to Superconductivity}, McGraw-Hill (1996)
\bibitem{kitt} C. Kittel, \textit{Introduction to Solid State Physics}, Wiley, New York (1986).
\bibitem{doll} R. Doll and M. N\"abauer, Phys. Rev. Lett. {\bf 7}, 51 (1961)
\bibitem{finn} D.K. Finnemore, T.F. Stromberg, and C.A. Swenson, Phys. Rev. {\bf 149} 231 (1966)
\bibitem{bc2}  M. Tinkham, Phys. Rev. {\bf 129}, 2413 (1963)
\bibitem{niob} H.A. Heupold and H.A. Boorse, Phys. Rev. {\bf 134}, 1322 (1964)
\bibitem{stefan} S. Stahl et al., Eur. Phys. J. D {\bf 32}, 139 (2005)
\bibitem{gold} J. Goldman and G. Gabrielse, Phys. Rev. A {\bf 81}, 052335 (2010)
\bibitem{pra} W. Quint, D. Moskovkin, V.M. Shabaev and M. Vogel, Phys. Rev. A {\bf 78}, 03251 (2008)
\bibitem{pra2} D. von Lindenfels \textit{et al.}, Phys. Rev. A  {\bf 87}, 023412 (2013)
\bibitem{inv1} L.S. Brown and G. Gabrielse, Phys. Rev. A {\bf 25}, 2423 (1982) 
\bibitem{inv2} G. Gabrielse, Phys. Rev. Lett. {\bf 102}, 172501 (2009)
\bibitem{sturm} S. Sturm et al., Phys. Rev. Lett. {\bf 107}, 143003 (2011)
\bibitem{gab84} G. Gabrielse and F.C. Macintosh, Int. J. Mass. Spec. Ion Proc. {\bf 57}, 1 (1984)
\bibitem{yu} J. Yu, M. Desaintfuscien and F. Plumelle, Appl. Phys. B {\bf 48}, 51 (1989)
\bibitem{win} D.F.A. Winters, M. Vogel, D.M. Segal and R.C. Thompson, J. Phys. B {\bf 39}, 3131 (2006)
\bibitem{spec} Z. Andjelkovic et al., Phys. Rev. A {\bf 87}, 033423 (2013)
\bibitem{stacey} F.D. Stacey and P.M. Davis, \textit{Physics of the Earth}, 4th edition, Cambridge University Press, Cambridge (2008)
\bibitem{sva} J. Bengtsson, J. Larsson, S. Svanberg and C.G. Wahlstr\"om, Phys. Rev. A {\bf 41}, 233 (1990) 

\end{thebibliography}
\end{document}